\documentclass[11pt]{article}

\usepackage{latexsym,amsmath,amsfonts,amssymb}

\usepackage[]{graphicx}

\def\be{\begin{equation}}
\def\ee{\end{equation}}
\def\tr{\ensuremath{\mathrm{Tr}}}

\begin{document}

\par\hfill CPHT-PC135.1206
\vskip 0.01in \hfill hep-th/0701059

\vspace{30pt}

\begin{center}
{\bf \Large A Note on the String Dual of $\mathcal{N}=1$ SQCD-like Theories}
\end{center}

\vspace{30pt}

\begin{center}
Roberto Casero\footnote{roberto.casero@cpht.polytechnique.fr}, Angel Paredes\footnote{angel.paredes@cpht.polytechnique.fr}

\vspace{30pt}

\textit{Centre de Physique Th\'eorique\\ Ecole Polytechnique\\ and UMR du CNRS 7644\\ 91128 Palaiseau\\France}

\end{center}

\vspace{30pt}


\begin{center}
{\bf Abstract}
\end{center}

\noindent In this note we describe how $\mathcal{N}=1$ SQCD-like theories with a large number of flavors can be given a dual description in terms of a string background containing a large number of additional D5-branes.
The dual geometries account for the backreaction of the additional  branes: they depend on the ratio between the number of flavors and colors of the gauge theory. 

\noindent This note is based on \cite{us}.  We present here also a new set of solutions, which are not  included in the original paper.

\section{Introduction}

The AdS/CFT correspondence has proved  as an incredibly powerful tool to study non-perturbative, strong coupling effects in gauge theories. Its application to supersymmetric and conformal field theories is rather well understood, and many checks and predictions have been carried out.  Approaching theories which are phenomenologically more interesting, though,  becomes much more complicated, and the string/gauge correspondence still has to display all of its power.

One of the primary steps in the construction of a string dual to QCD-like theories is the addition of flavor degrees of freedom to the general picture. This can be done by adding a set of additional D-branes to the string background dual to a theory of (super)glue plus some well defined UV completion which is determined by the initial brane setup \cite{kk}.

When the number $N_f$ of these flavor branes  is much smaller than the number $N_c$ of colors  of the gauge theory, corrections to the background metric and fluxes due to the presence of the flavor branes are of order $N_f/N_c$ and can therefore be neglected. This probe approximation allows for the accounting and description of several non-trivial aspects of the dynamics of flavors.

However, some very interesting physics of QCD-like theories appears only when the ratio $N_f/N_c$ is kept finite, and cannot be explored in the probe approximation of the gauge/string correspondence. For this reason it seems very important to find a way to  describe backgrounds where the effect and backreaction of flavor branes is taken into account.

In this note we report the progress in this direction made in \cite{us}, where such a problem was studied for a bunch  of flavor D5-branes added to the wrapped brane background \cite{MN} dual to $\mathcal{N}=1$ SYM. This results in a background dual to a set of $\mathcal{N}=1$ SQCD-like theories with a large number $N_f$ of flavors.\footnote{The procedure of \cite{us} has been applied in \cite{Benini:2006hh} starting from  a class of $\mathcal{N}=1$ superconformal field theories and adding many flavors. One interesting point of \cite{Benini:2006hh}  is that the original dual color theory is precisely known, even in its far UV region. This allows for a more detailed understanding of some aspects of the present approach.} We will moreover present a new set of solutions \cite{us2} which were not included in the original paper \cite{us}.

\section{The dual to $\mathcal{N}=1$ super Yang-Mills}\label{sect: MN}

Let us start by introducing the gravity solution derived in \cite{MN}, which is dual to $\mathcal{N}=1$ SYM. This solution is the background from which our study of the backreaction of flavor branes starts, and moreover this section will allow us to introduce notation and conventions we will use throughout this note. 

This solution corresponds to D5-branes wrapping a non-trivial two-cycle inside the resolved conifold, and undergoing a geometric transition. The solution in the Einstein frame reads
\be\label{metricaa}
ds^2_{10}\,=\,\alpha' g_s N_c \,e^{\frac{\phi}{ 2}} \Big[\,
\frac{1}{\alpha' g_s N_c }dx^2_{1,3}\,+\,
dr^2+\,e^{2h}\,\big(\,d\theta^2+\sin^2\theta 
d\varphi^2\,\big)\,\,+\,\frac{1}{4}\,(\tilde{\omega}_i-A^i)^2\,\Big]
\ee
where $\phi$ is the dilaton. Here $\psi$ has periodicity $4\pi$. The angles $\theta\in [0,\pi]$ and $\varphi\in [0,2\pi)$ parametrize a two-sphere. 
This sphere is fibered in the ten-dimensional metric by the one-forms
$A^i$ $(i=1,2,3)$.
They are given in terms of a function
$a(r)$ and the angles $(\theta,\varphi)$ as follows:
\be
A^1\,=\,-a(r) d\theta \qquad \quad A^2\,=\,a(r) \sin\theta d\varphi \qquad\quad A^3\,=\,- \cos\theta d\varphi
\label{oneform}
\ee
The $\tilde{\omega}_i$ one-forms are defined as
\be\label{su2}
\tilde{\omega}_1= \cos\psi \,d\tilde\theta+\sin\psi\sin\tilde\theta \,d\tilde\varphi\qquad \quad \tilde{\omega}_2=-\sin\psi \,d\tilde\theta+\cos\psi\sin\tilde\theta \,d\tilde\varphi \qquad\quad \tilde{\omega}_3=d\psi+\cos\tilde\theta\, d\tilde\varphi
\ee
The geometry in (\ref{metricaa}) preserves four supercharges and is 
non-singular  
when the functions $a(r)$, $h(r)$ and the dilaton $\phi(r)$ are
\be\label{MNsol}
a(r)=\frac{2r}{ \sinh 2r}\qquad\quad e^{2h}=r\coth 2r\,-\,\frac{r^2}{ \sinh^2 2r}-\frac{1}{ 4}\qquad\quad e^{-2\phi}=e^{-2\phi_0}\frac{2e^h}{ \sinh 2r}
\ee
where $\phi_0$ is the value of the dilaton at $r=0$. 
The  type IIB supergravity solution includes a
RR three-form $F_{(3)}$ that is given by
\be\label{RRthreeform}
\frac{1}{g_s \alpha' N_c} F_{(3)}\,=\,-\frac{1}{4}\,\big(\,\tilde{\omega}_1-A^1\,\big)\wedge
\big(\,\tilde{\omega}_2-A^2\,\big)\wedge \big(\,\tilde{\omega}_3-A^3\,\big)\,+\,\frac{1}{4}\,\,
\sum_a\,F^a\wedge \big(\,\tilde{\omega}_a-A^a\,\big)
\ee
where $F^a$ is the field strength of the $SU(2)$ gauge field $A^a$, defined as $F^a\,=\,dA^a\,+\,\frac{1}{2}\epsilon_{abc}\,A^b\wedge A^c$. The different components of $F^a$ read                                                                               
\be
F^1=-a'\,dr\wedge d\theta\qquad \quad F^2=a'\sin\theta\, dr\wedge d\varphi\qquad \quad F^3=(1-a^2)\sin\theta \,d\theta\wedge d\varphi
\ee
where prime denotes derivative with respect to $r$. 

This solution is invariant under $SU(2)\times SU(2) \times U(1)_R$ transformations.

\section{Adding Flavor Branes}\label{sect: adding flavors}

In this section we will add a large number $N_f$ of flavor branes to the background \cite{MN} presented in the previous section. Massless flavors can be introduced by means of  D5-branes embedded along the  {\it cylinder solutions} that were found in \cite{NPR}: they are extended along the 4-dimensional Minkowski directions $x^0$, $x^1$, $x^2$, $x^3$ and the radius $r$, and wrap the angle $\psi$. These branes preserve the same supersymmetries as the background, and the gauge theory $U(1)_R$ symmetry \cite{NPR} which is associated to shifts of the $\psi$ angle.

When we go beyond the probe approximation, the effect of the branes on the background cannot be neglected. To take it into account  we follow the procedure introduced in \cite{KM} in the context of non-critical strings, and add an open string sector to the gravity action
\be
S=S_{grav}+S_{flavor}
\ee
where $S_{grav}$ in the Einstein frame reads
\be\label{gravaction}
S_{grav}=\frac{1}{2\kappa_{(10)}^2}
\int d^{10}x \sqrt{-g} \left[R-\frac12 (\partial_\mu \phi)
(\partial^\mu \phi)-\frac{1}{12}e^{\phi}F_{(3)}^2\right]
\ee
and $S_{flavor}$ is the Dirac + Wess-Zumino action for the $N_f$ D5-flavor branes (in the Einstein frame)
\be
\label{sflavor}
S_{flavor}=T_5 \sum^{N_f} \left(
-\int_{{\cal M}_6} d^{6} x\; e^{\frac{\phi}{2}}\sqrt{-\hat g_{(6)}}
+ \int_{{\cal M}_6} P[C_{(6)}] \right)
\ee
Here the integrals are taken over the six-dimensional worldvolume $\mathcal{M}_6$ of the flavor branes, and $P[C_{(6)}]$ and $\hat{g}_{(6)}$  stand respectively for the pull-back on $\mathcal{M}_6$ of the RR bulk 6-form potential, and for the determinant of the pull-back of the metric. The worldvolume gauge field can  be put to zero consistently.

In a generic configuration the flavor D5-branes can be arbitrarily distributed on the transverse directions $\theta$, $\varphi$, $\tilde{\theta}$, $\tilde{\varphi}$.  In the field theory this corresponds to introducing the flavors in such a way that breaks most of the global symmetries of the lagrangian. On the gravity side, therefore, most of the isometries of the background would be lost, making it practically impossible to find a solution to the equations of motion of the system. To avoid this problem, it is desirable to find a coupling of the flavor degrees of freedom that preserves the original global symmetries of the field theory. This is achieved by homogeneously smearing  the $N_f\rightarrow \infty$ flavor branes  along the two transverse $S^2$'s parameterized by $\theta$, $\varphi$ and $\tilde{\theta}$, $\tilde{\varphi}$ \cite{us}. This configuration is still $SU(2)\times SU(2) \times U(1)_R$ symmetric, as required.

For the metric we can therefore take the general ansatz
\be\label{nonabmetric}
\begin{split}
ds^2 =& \;e^{2 f(r)} \Big[dx_{1,3}^2 + dr^2 + e^{2 h(r)} (d\theta^2 + \sin^2\theta d\varphi^2) +\\
&+\frac{e^{2 g(r)}}{4} \left((\tilde{\omega}_1+a(r)d\theta)^2 + (\tilde{\omega}_2-a(r)\sin\theta d\varphi)^2\right) + \frac{e^{2 k(r)}}{4} (\tilde{\omega}_3 + \cos\theta d\varphi)^2\Big] 
\end{split}
\ee
Notice that compared to (\ref{metricaa}), we took $\alpha' g_s =1$, 
while $N_c$ has been absorbed in $e^{2h}$, $e^{2g}$, $e^{2k}$. The background also contains a non-trivial dilaton $\phi(r)=4f(r)$ and a 3-form RR field strength. After we substitute (\ref{nonabmetric}) into the action for the open string sector (\ref{sflavor}) and we take into account the effect of the smearing of the flavor branes, $S_{flav}$ reads
\be\label{S smeared}
S_{flav}=\frac{T_5\,N_f}{(4\pi)^2}\left(-\int d^{10}x  \sin \theta \sin\tilde{\theta}\,e^{\frac{\phi}{2}}\sqrt{-\hat g_{(6)}}\;+\int \mathrm{Vol}(\mathcal{Y}_4)\wedge C_{(6)}\right)
\ee
where we have defined $\mathrm{Vol}(\mathcal{Y}_4)=\sin\theta\sin\tilde{\theta}\,d\theta\wedge d\varphi\wedge d\tilde{\theta}\wedge d\tilde{\varphi}$ and the integrals span the whole space-time now.

The WZ term of the flavor brane action, that is the second term on the r.h.s. of (\ref{S smeared}), depends neither on the metric nor on the dilaton. Therefore, it does not enter the Einstein equations. However, it contributes to the equations of motion of the 6-form $C_{(6)}$, which in the unflavored case was $d*F_{(7)}\equiv dF_{(3)}=0$. Now, the smeared flavor branes act as a uniform source for the RR form and we have
\be\label{newdF}
dF_{(3)} = \frac14 N_f \,\mathrm{Vol}({\cal Y}_4) =
\frac14 N_f \sin \theta \sin \tilde \theta\,
d\theta \wedge d\varphi \wedge d\tilde\theta \wedge d\tilde\varphi
\ee
We can solve this equation by writing $F_{(3)}$ as the sum of two terms. The first one is fixed by the symmetries of the background and satisfies\footnote{Notice that by setting $b(r)=a(r)$, the following equation (\ref{F3ans}) is exactly the unflavored three-form (\ref{RRthreeform}).} $dF_{(3)}=0$
\be\label{F3ans}
\begin{split}
F_{(3)}=\;\frac{N_c}{4}\Big[&-(\tilde{\omega}_1+b(r) d\theta)\wedge
(\tilde{\omega}_2-b(r) \sin\theta d\varphi)\wedge
(\tilde{\omega}_3 + \cos\theta d\varphi)+ \\
&+b'dr \wedge (-d\theta \wedge \tilde{\omega}_1  + \sin\theta\, d\varphi \wedge 
\tilde{\omega}_2) + (1-b(r)^2) \sin\theta \,d\theta\wedge d\varphi \wedge
\tilde{\omega}_3\Big]
\end{split}
\ee
where $b(r)$ is a function to be determined. The second term accounts for the  Bianchi identity (\ref{newdF})
\be\label{flav3form}
F_{(3)}^{flavor} = 
-\frac{N_f}{4} 
\sin\theta \,d\theta \wedge d\varphi \wedge ( d\psi + \cos 
\tilde \theta\, d\tilde\varphi)
\ee

At this point we have all we need to derive the equations of motion for our system. Since we are looking for a background with four supersymmetries, the most convenient procedure is to plug the ansatz (\ref{nonabmetric}, \ref{F3ans}, \ref{flav3form}) into the equations for the supersymmetric transformations of the dilatino and gravitino \cite{NPR,us}, which provides us with a set of first-order BPS equations. As shown in \cite{us}, the system can be partially solved to give ($\rho$ is defined by $e^{-k(r)}dr=d\rho$, and $x\equiv \frac{N_f}{N_c}$)
\be\label{solved}
b=\frac{\left(2-x\right) \rho }
{\sinh (2\rho)} \qquad \quad  e^{2g}= \frac{N_c}{2}\, \frac{2b 
\cosh 2\rho -2+x}{a \cosh2\rho -1}\qquad \quad e^{2h}=\frac{e^{2g}}{4} (2 a \cosh(2\rho) -1 - a^2)
\ee
plus two coupled first order eqs for $a(\rho), k(\rho)$ (here $\tilde{k}\equiv k-\frac{1}{2}\log N_c$)
\be\label{eqs}
\begin{split}
\partial_\rho a=\;&\frac{2}{2\rho \coth 2\rho -1}
\left(-\frac{2e^{2\tilde{k}}-x}{2-x} \frac{(a\cosh 2\rho -1)^2}
{\sinh 2\rho}+a^2 \sinh 2\rho-2a\rho\right)\\
\partial_\rho \tilde{k}=\;& \frac{2}{(2\rho \coth 2\rho -1)
(1-2a\cosh 2\rho+a^2)}\left(  \frac{2e^{2\tilde{k}}+x}{2-x} a\sinh 
2\rho (a\cosh 2\rho -1)  +\right. \\
&\left.+2\rho(a^2\frac
{\sinh 2\rho}{2\rho}\cosh 2\rho-2a\cosh 2\rho +1)  \right)
\end{split}
\ee
 and  a differential equation for $f =\frac{\phi}{4}$ that can be integrated once  $a(\rho)$ is known
\be\label{feqflav}
\begin{split}
\partial_\rho f =\;&\frac{(-1+a \cosh 2 \rho) \sinh^{-2} (2 \rho)  }{4 (1+a^2-2 a \cosh 2 \rho) (-1+2 \rho  \coth 2 \rho )}\Big[-4 \rho +\sinh 4 \rho + 4a \rho \cosh 2\rho+ \\
& -2a \sinh 2\rho - \frac{4}{(2-x)} a (\sinh 2\rho)^3 \Big]
\end{split}
\ee
It is possible to check (using {\it Mathematica}) that the solutions to the BPS conditions (\ref{solved}, \ref{eqs}, \ref{feqflav})  solve the full system of second order Einstein equations and the flux equation $d*F_{(3)}=0$ \cite{us}.

\section{Solutions of the backreacted system}\label{sect: solutions}
It proves very difficult to solve equations (\ref{eqs}, \ref{feqflav}) analytically for a generic value of $N_f/N_c$. Nonetheless, it is still possible to study the problem numerically \cite{us}. The way to proceed is the following: first of all the equations are solved asymptotically by a power series expansion, near $\rho\rightarrow\infty$ (the UV of the dual gauge theory) and near $\rho=0$ (the IR of the dual theory). Then numerical integration is performed to find a solution that interpolates between the two asymptotic regions and behaves in the desired way in the IR and UV. For many purposes, this procedure proves as good as finding an explicit exact solution. For all details about the asymptotic expansion and the numerical integration of equations (\ref{eqs}, \ref{feqflav}) we refer the reader to \cite{us}. Here we will only report about the solutions.

\begin{figure}[!htb] 
   \includegraphics[width=0.4\textwidth]{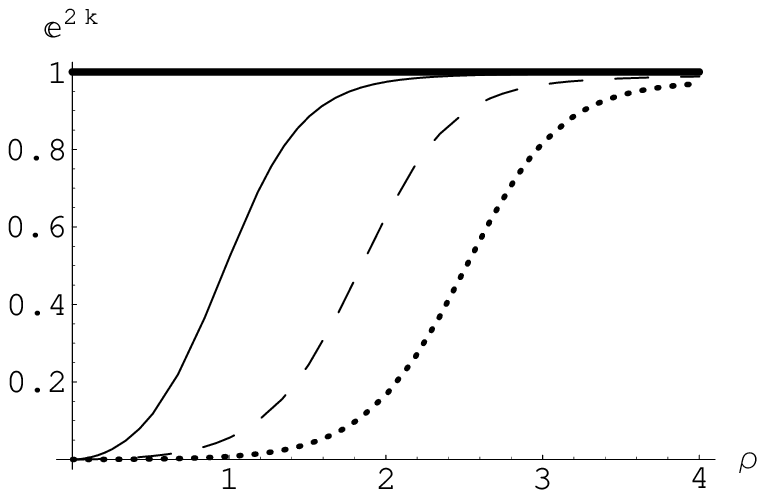} \hfil
   \includegraphics[width=0.4\textwidth]{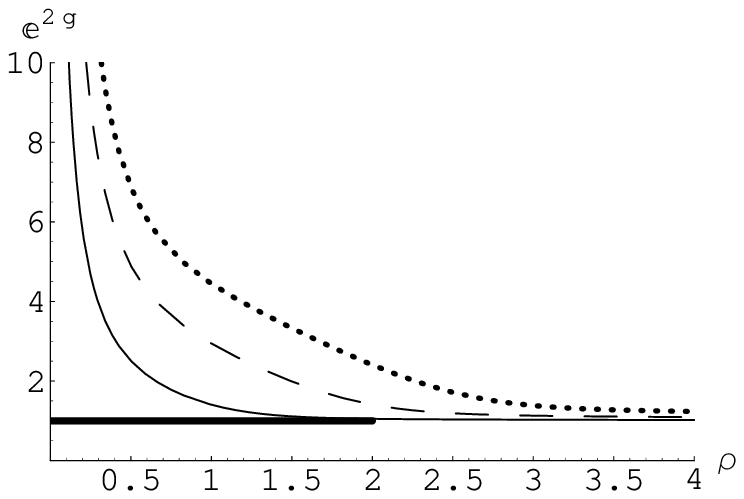} \\
   \includegraphics[width=0.4\textwidth]{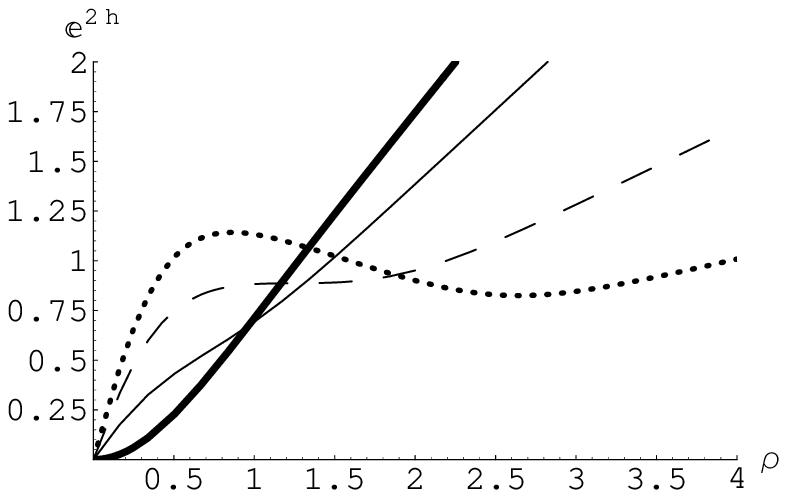} \hfil
   \includegraphics[width=0.4\textwidth]{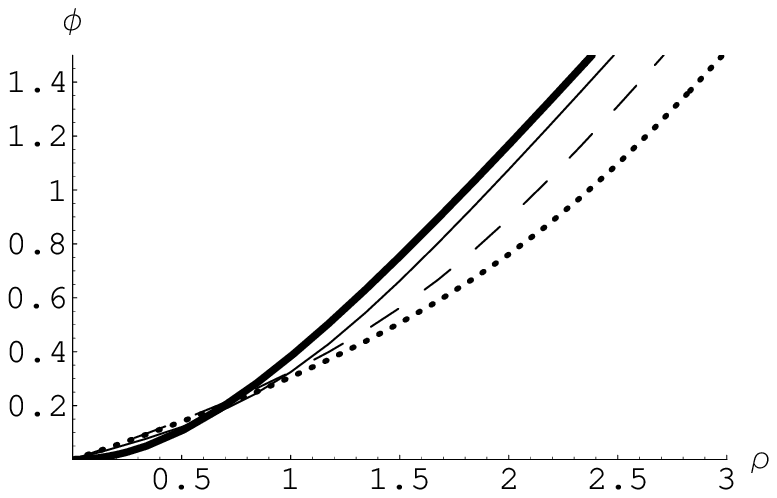} \\
   \includegraphics[width=0.4\textwidth]{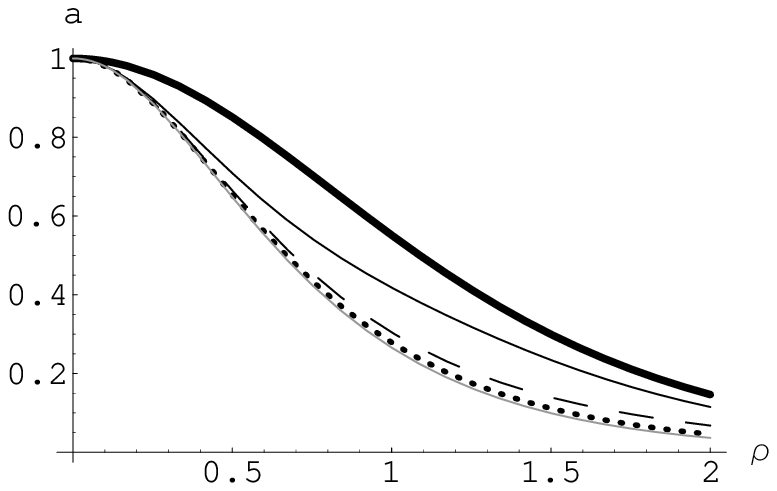} \hfil
   \includegraphics[width=0.4\textwidth]{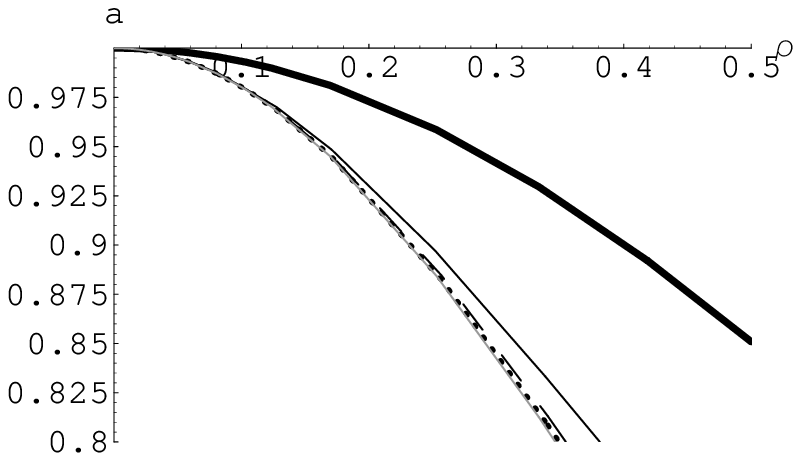} 
   \caption{Some functions \cite{us} of the flavored solutions for
   $N_f = 0.5 N_c$ (thin solid lines), $N_f=1.2 N_c$ (dashed lines),
   $N_f = 1.6 N_c$ (dotted lines). For comparison, we also plot
   (thick solid lines) the usual unflavored solution. The graphs
   are $e^{2k}$, $e^{2g}$, $e^{2h}$, $\phi$, $a$ and a zoom
   of  the plot for $a$ near $\rho=0$. In the figures for $a$ we have
   also plotted with a light solid line $a=\frac{1}{\cosh (2\rho)}$.}
   \label{flavoredgraphs}
\end{figure}

\begin{figure}[!htb] 
   \includegraphics[width=0.4\textwidth]{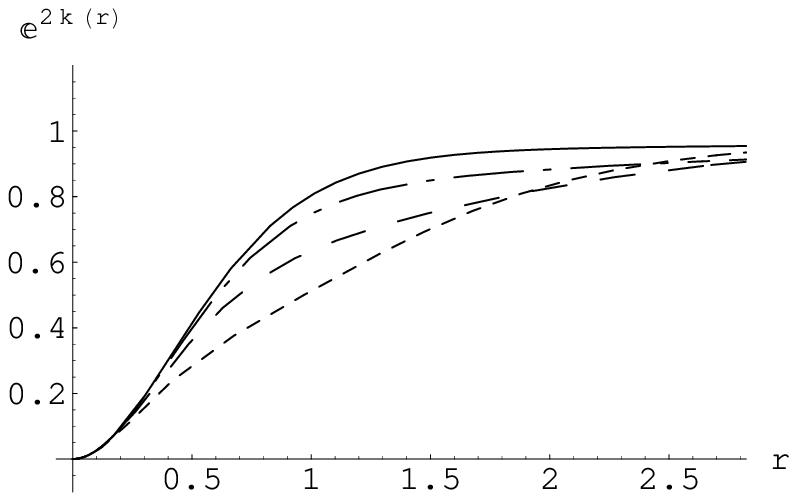} \hfil
   \includegraphics[width=0.4\textwidth]{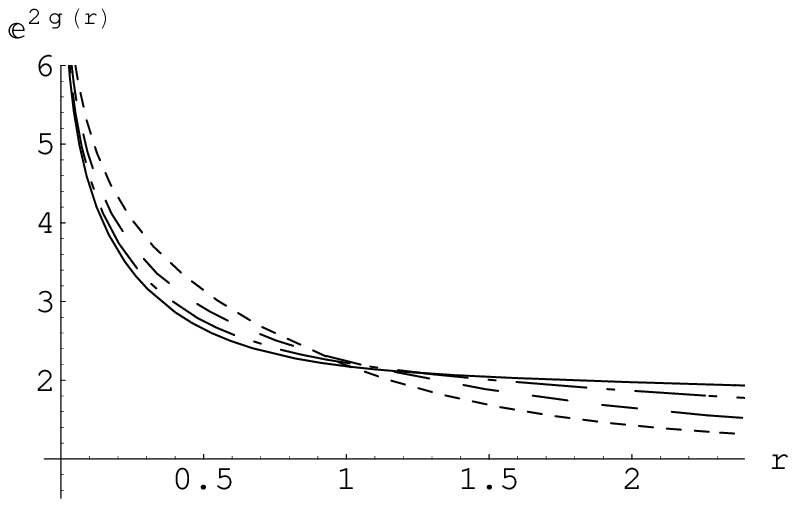} \\\vskip 2pt
   \includegraphics[width=0.4\textwidth]{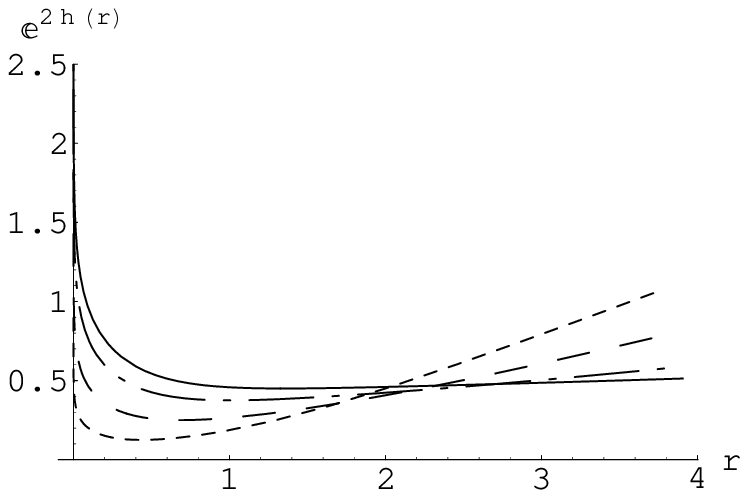} \hfil
   \includegraphics[width=0.4\textwidth]{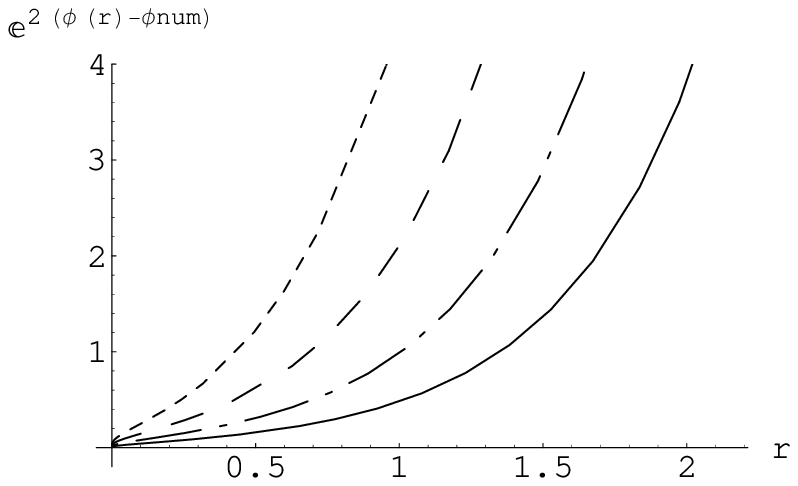} 
   \caption{Some functions \cite{us2} of the flavored solutions with $a(r)=b(r)=0$, for
   $N_f = 1.25 N_c$ (short-dashed lines), $N_f=1.5 N_c$ (long-dashed lines),
   $N_f = 1.75 N_c$ (dash-dotted lines) and $N_f=1.9 N_c$
   (solid lines). In the last graph, $\phi_{num}$ is an integration constant. Notice that  $e^{2\phi}$ vanishes at the origin.}   \label{abelianflavoredgraphs}
\end{figure}

The numerical integration results are shown in figures \ref{flavoredgraphs} and \ref{abelianflavoredgraphs}. The first thing that should be noted is that all solutions present a singularity at the origin. Nonetheless, these singularities are of a good kind, as classified in \cite{MN1}, that is in the Einstein frame $g_{tt}$ is bounded from above: an excitation 
travelling towards the origin (that is a low energy object on the gauge 
theory dual) will have 
an energy as measured by an inertial gauge theory observer given  by 
$E=\sqrt{|g_{tt}|} E_0$ (with $E_0$ the proper energy of the object);
if $g_{tt}$ is bounded from above, the energy of the object remains finite from the inertial observer point of view, which is physically acceptable.

The classification of good and bad singularities allows us to distinguish among three different regions of the field theory parameter space: $N_f<N_c$, $N_c<N_f<2N_c$, and $N_f=2N_c$. The case $N_f>2N_c$ was not considered in \cite{us}.

For $N_f<N_c$ we have a single kind of solutions, with $a(r)\neq 0$, as shown in figure \ref{flavoredgraphs}. Setting $a(r)= 0$ in this case, we could only find backgrounds with unacceptable singularities at the origin \cite{us2}. For $N_c<N_f<2N_c$, on the contrary, alongside solutions with $a(r)\neq 0$ (figure \ref{flavoredgraphs}), we also have solutions with $a(r)= b(r)=0$ \cite{us2} which present a good singularity at the origin (figure \ref{abelianflavoredgraphs}). These two different kinds of solutions might represent two different phases of the flavored gauge theory, with the new one corresponding to $a(r)= b(r)=0$ only present for $N_f>N_c$. Finally for $N_f=2N_c$ an explicit solution to the differential equations above can be found, which gives a background  dual to a field theory which has many features of a conformal theory \cite{us}.

\section{Gauge Theory Aspects and Predictions of the Solution}

Once the backgrounds corresponding to the addition of many flavor branes to a solution which is dual to some color gauge theory are found, it is of the foremost importance to identify the precise field theory which is dual to the backreacted backgrounds. The problem is made more complicated by the fact that in the supergravity approximation the original unflavored background is not simply dual to pure (supersymmetric) Yang-Mills, but includes additional states transforming in the adjoint representation of the gauge group and having a characteristic scale of the same order as the strong coupling scale of the gauge theory. Being in the adjoint representation, these additional fields allow for various possible couplings to the flavor degrees of freedom, in addition to the minimal gauge coupling.

For the backgrounds presented in section \ref{sect: solutions}, this problem has been addressed in \cite{us}. In that paper it is argued that the coupling between the additional adjoint fields and the fundamental matter is reminiscent of the coupling of the adjoint chiral superfield in the $\mathcal{N}=2$ vector multiplet to a fundamental hypermultiplet. After integrating out the additional adjoint fields, the most relevant contribution to the superpotential for the fundamental fields can be written schematically as
\be\label{superp}
\mathcal{W}\sim \tr(\tilde{Q}Q\tilde{Q}Q)+\tr(\tilde{Q}Q)^2
\ee
which might possibly be completed by operators due to higher order couplings of the KK adjoint fields.

In the next subsections we will report some non-trivial computations performed on the backreacted backgrounds of section \ref{sect: solutions}, which faithfully reproduce expected properties of the field theory in (\ref{superp}).

\subsection{Wilson Loop and Color Charge Screening}
Massless flavors are expected to drastically change the behavior of Wilson loops. Whereas for small separations of the heavy quark pair, we expect a behavior which is very similar to a confining gauge theory, when the distance between the quark and antiquark is of the same order as the mass of the lightest mesons, it is more favorable for the vacuum to produce an additional pair of massless quarks, which combine to the two heavy quarks to give two mesons. Being the color charge of the original quarks screened by the newly created pair, the two mesons can be arbitrarily separated without any effort.

The computation of the Wilson loop in the dual background can be reproduced by studying an open string with the endpoints attached to the boundary of space and stretching down into the bulk \cite{Wilsonloop}. In the case of the solutions in figure \ref{flavoredgraphs} which have $a(r)\neq 0$, there exists a maximum allowed distance between the endpoints of such strings: longer strings are not solutions of the Nambu-Goto action.  The relation between the energy of the allowed strings and the distance between their endpoints  reproduces precisely the color charge screening behavior we described above \cite{us}. In the case of the solutions in figure \ref{abelianflavoredgraphs} (which exist only for $N_f>N_c$), instead, the exponential of the dilaton goes to zero at the origin and the QCD string is tensionless in the IR. This might suggest that the $a=0$ phase of the flavored gauge theory is not confining.

\subsection{UV Beta function and $U(1)_R$ breaking}

The UV values of the gauge coupling and theta angle can be evaluated in the string picture by wrapping a euclidean D1-brane on a supersymmetric two-cycle of the internal geometry, which in the dual field theory corresponds to an instanton. The action of this brane reads
\be\label{actiond1a}
S=  T_1\int d^2 z e^{-\phi}\sqrt{det g} + T_1\ i \int C_{(2)}=
\frac{8\pi^2}{g_{sqcd}^2} + i\theta_{sqcd}
\ee
By picking the supersymmetric two-cycle $\theta=\tilde{\theta}$, $\tilde{\varphi}=2\pi-\varphi$, $\psi= (2n+1)\pi$, $\rho\rightarrow \infty$  \cite{DiVecchia:2002ks},
and using the definition of $C_{(2)}$ given in \cite{us}, the expression (\ref{actiond1a}) gives
\be\label{gtheta}
\theta_{sqcd}= \frac12 (2N_c- N_f)(\psi-\psi_0) \hfil 
\frac{4\pi^2}{g_{sqcd}^2}= e^{2h} + \frac{e^{2g}}{4}(a-1)^2
\ee
We can now use these expressions to evaluate the UV beta function of the dual gauge theory and identify the $U(1)_R$ breaking pattern. For the beta function, by using the solutions of section \ref{sect: solutions} it is found \cite{us}
\be
\beta=\frac{d g_{sqcd}}{d\log\frac{\mu}{\Lambda}}= -\frac{3 g_{sqcd}^3}
{32 \pi^2}(2 N_c 
- N_f)+\ldots
\ee
which matches the behavior of softly broken $\mathcal{N}=2$ SQCD, that is the UV completion of the superpotential~(\ref{superp}). 

The expression for the theta angle $\theta_{sqcd}$ (\ref{gtheta}) shows that arbitrary rotations of $\psi$, which correspond to $U(1)_R$ transformations, are not symmetries of the field theory, the theta angle being shifted by a non-integer multiple of $2\pi$. It turns out then, that only shifts $\psi\rightarrow \psi+\frac{4\kappa\pi}{2N_c-N_f}$ with $\kappa=1,2,\ldots, 2N_c-N_f$ (notice that $\psi$ has $4\pi$ periodicity) leave $\theta_{sqcd}$ unchanged. Quantum effects, then, select a discrete subgroup $\mathbb{Z}_{2N_c-N_f}$ out of the $U(1)_R$ $R$-symmetry, which is a realization of the anomalous breaking of $R$-symmetry in $\mathcal{N}=1$ supersymmetric theories. The rank of the surviving group matches the expected result for a theory with $N_f$ flavors. For an equivalent derivation of the $R$-symmetry anomaly in a different string/gauge correspondence setup see \cite{Klebanov:2002gr}.

The non-anomalous, discrete, $R$-symmetry group is further broken by gaugino condensation, which is reflected in the geometry being invariant under only a $\mathbb{Z}_2$ subgroup of $\mathbb{Z}_{2N_c-N_f}$ when  either $a$ or $b$ or both are turned on \cite{Apreda:2001qb}. Notice that these functions are non-zero only in the IR region of the background, which is consistent with a spontaneous breaking of the $R$-symmetry.

\subsection{Seiberg Duality}

One of the most amazing features of $\mathcal{N}=1$ theories with matter in the fundamental representation  is Seiberg-duality~\cite{Seiberg:1994pq}.  In the present construction, Seiberg 
duality appears in a remarkably simple and elegant way: it corresponds to switching between two alternative ways of writing the background. The central point which the realization of Seiberg duality in the model of section \ref{sect: adding flavors} is based on, is that the internal space of 
our background has $S^2\times S^3$ topology, but the choice to use either the 
space spanned by the pair $(\theta, \varphi)$ or 
by  $(\tilde{\theta}, \tilde{\varphi})$ as the $S^2$ base  of the $S^3$ is  (at least at first sight)  arbitrary. 

Remarkably, one way of writing the background corresponds to a field theory with $N_c$ colors and $N_f$ flavors, whereas the other one represents a theory with $\bar{N}_c=N_f-N_c$ colors and again $N_f$ flavors \cite{us}.  Since the structure of the background remains the same, we expect the two Seiberg-dual theories to have the same lagrangian. All these features match exactly what we expect from Seiberg duality applied to $\mathcal{N}=1$ SQCD deformed by a quartic superpotential for the quark superfields \cite{strassler}. Moreover, when the two geometries are compared in the IR, one finds that the behavior of the two backgrounds matches up to the first few orders in the small radius expansion \cite{us}. This is another essential feature of Seiberg duality, which states the identity of the IR dynamics of the pair of dual theories.

\subsubsection*{Acknowledgments}

It is a pleasure to thank Carlos N\'u\~nez for many useful and illuminating discussions and for our collaboration in  \cite{us}. RC would also like to thank  the organizers of the workshop of the Marie Curie Research Training Network ``Constituents, Fundamental Forces and Symmetries of the Universe'', Naples, October 2006, where this work was presented.  RC and AP are supported by European Commission Marie Curie Postdoctoral Fellowships, under contracts MEIF-CT-2005-024710 and MEIF-CT-2005-023373, respectively. The work of  RC and AP  is also partially supported by INTAS grant 03-51-6346, CNRS PICS \# 2530, RTN contracts MRTN-CT-2004-005104 and MRTN-CT-2004-503369 and by a European Union Excellence Grant MEXT-CT-2003-509661.

\end{document}